\documentclass{gGAF2e}
\usepackage{amsmath, amssymb, bm}

\usepackage{multiequations}

\newcommand*{\be}{\begin{equation}}
\newcommand*{\ee}{\end{equation}}
\newcommand*{\bea}{\begin{eqnarray}}
\newcommand*{\eea}{\end{array}}
\newcommand*{\bal}{\begin{align}}
\newcommand*{\eal}{\end{align}}
\newcommand*{\bme}{\begin{multiequations}}
\newcommand*{\eme}{\end{multiequations}}
\newcommand*{\se}{\singleequation}

\newcommand*{\qe}{\quadequation}

\renewcommand*{\Omega}{\varOmega}

\begin{document}

\title{Local analysis of the magnetic instability in rotating magneto-hydrodynamics with the short-wavelength approximation}
\author{Xing Wei\thanks{Email: xingwei@princeton.edu}\\Institute of Geophysics, University of G\"ottingen \\ Friedrich-Hund-Platz 1, G\"ottingen 37077, Germany
\\
\vspace{6pt}\received{Received 1 February 2013; in final form 5 August 2013; first published online ????} }
\maketitle

\begin{abstract}
We investigate analytically the magnetic instability in a rotating and electrically conducting fluid induced by an imposed magnetic field with its associated electric current. The short-wavelength approximation is used in the linear stability analysis, i.e. the length scale of the imposed field is much larger than the wavelength of perturbations. The dispersion relationship is derived and then simplified to give the criteria for the onset of the magnetic instability in three cases of imposed field, namely the axial dependence, the radial dependence and the mixed case. The orientation of rotation, imposed field and imposed current is important for this instability.\\

\end{abstract}

\section{Introduction}
Magnetic field is ubiquitous in the universe. Dynamo action is believed to generate magnetic field through the electromagnetic induction effect, namely the motion of conducting fluid shears and twists the field lines to create the new field lines to offset the magnetic field which diffuses away. In self-sustained dynamo action, fluid flow and magnetic field are nonlinearly coupled such that the Lorentz force suppresses the growth of magnetic field by damping the amplitude and changing the spatial structure of fluid flow. However, most recently, another role of Lorentz force was discovered in the nonlinear subcritical dynamo, i.e. the Lorentz force facilitates the dynamo action \citep{jones}. This might be caused by the magnetic instability, i.e. the instability in the magneto-hydrodynamic (MHD) flow induced by an externally imposed magnetic field with its associated electric current.

In the geophysical context where incompressible fluid is usually considered, the magnetic instability in rotating MHD was initially studied in cylindrical annulus for the ideal MHD \citep{acheson} and then thoroughly developed for the resistive MHD \citep{fearn1,fearn2,fearn3,fearn4}. The magnetic instability was also studied in the magnetoconvection problem (convection with an externally imposed magnetic field) in the spherical shell geometry which is more relevant to the Earth's core \citep{zhang1,zhang2,zhang3,proctor}. In the astrophysical context where compressibility takes its effect, the magnetic instability was studied in cylindrical geometry \citep{tayler}. In the plasma physical context, the instability caused by both a shear flow and a shear field was investigated in the infinite space \citep{ofman}. Most recently, the magnetic instability was theoretically studied in Taylor-Couette setup \citep{ruediger1,ruediger2} and has already been identified in a liquid metal experiment \citep{ruediger3}. In this paper we study a much simpler situation, i.e. the instability due to a weak electric current with the absence of shear flow in the local Cartesian geometry. In section \ref{B-S} we give the basic state, in sections \ref{Bx-2}, \ref{Bx-3} and \ref{Bx-23} we formulate the linear stability analysis and simplify the stability criteria in three cases, i.e. the imposed field varies with radial coordinate or axial coordinate or both, and in section \ref{Summary} we make a brief summary of the results.

\section{Basic state \label{B-S}}
The Navier-Stokes (N-S) equation for incompressible MHD in a rotating frame is
\begin{equation}
\label{Navier-Stokes}
\frac{\partial\bm U}{\partial t}+\bm U\bm\cdot\bm\nabla\bm U=-\frac{1}{\rho}\bm\nabla P+\nu\nabla^2\bm U+2\bm U\times\bm\Omega+\frac{1}{\rho}\bm J\times\bm B,
\end{equation}
where $\bm U$ is the fluid velocity, $\rho$ is the fluid density, $P$ is the reduced fluid pressure involving the curl-free centrifugal force, $\nu$ is the fluid viscosity, $\bm\Omega$ is  the rotation vector, $\bm B$ is the magnetic field and $\bm J$ is the electric current. The magnetic induction equation is
\begin{equation}
\label{magnetic-induction}
\frac{\partial\bm B}{\partial t}=\bm\nabla\times\left(\bm U\times\bm B\right)+\eta\nabla^2\bm B,
\end{equation}
where $\eta$ is the magnetic diffusivity.

We decompose the variables into their background and perturbation parts,
\bme
\label{decomposition}
\begin{equation}
\qe
\bm U=\bm U_0+\bm u, \hspace{10mm} P=P_0+p, \hspace{10mm} \bm B=\bm B_0+\bm b, \hspace{10mm} \bm J=\bm J_0+\bm j,
\end{equation}
\eme
and study local stability relative to the Cartesian co-ordinate system $(x_1, x_2, x_3)$, with unit vectors $(\bm e_1, \bm e_2, \bm e_3)$. The rotation vector $\bm\Omega$ is assumed to be in the direction of $\bm e_3$, the azimuthal direction is locally $\bm e_1$ and the radial direction is $\bm e_2$. In this local coordinate system the effects of curvature are neglected. Though curvature effects might be important for dynamo action (say, Rossby-wave induced $\alpha$ effect), they will not be considered in our simplified model. There is no motion in the background state, i.e. $\bm U_0=\bm 0$, other than the rigid rotation
of our reference frame. Consequently shear flow instabilities are absent and we can focus on the magnetic instability. In planetary or stellar interior the large-scale shear arising from convection creates a large-scale steady axisymmetric azimuthal field (the so-called $\Omega$ effect), and this large-scale field is assumed to be the background field $\bm B_0=B_0(x_2,x_3)\bm e_1$. In the basic state, the magnetic induction equation is reduced to $\nabla^2\bm B_0=\bm 0$. 

If $B_0$ depends only on the axial coordinate $x_3$, i.e. $B_0=B_0(x_3)$ then the Laplacian equation of $\bm B_0$ requires that its associated electric current $\bm J_0$ should be uniform in the radial direction $\bm e_2$. The basic state,
which we consider in section \ref{Bx-2}, is
\bme
\label{background1}
\begin{equation}
\qe
\bm U_0=\bm 0, \hspace{10mm} \bm\Omega=\Omega\bm e_3, \hspace{10mm} \bm B_0=B_0(x_3)\bm e_1, \hspace{10mm} \bm J_0=J_0\bm e_2,
\end{equation}
and the background field and the background current are related through
\begin{equation}
\se
\frac{\mathrm d B_0}{\mathrm d x_3}=\mu J_0,
\end{equation}
\eme
where $\mu$ is the magnetic permeability. 

On the other hand, if $B_0$ depends only on the radial coordinate $x_2$, i.e. $B_0=B_0(x_2)$, which might be more interesting to astrophysicists \citep{acheson,tayler}, then its associated electric current $\bm J_0$ is also uniform but in the axial direction $\bm e_3$. Then the basic state,
which we consider in section \ref{Bx-3}, is 
\bme
\label{background2}
\begin{equation}
\qe
\bm U_0=\bm 0, \hspace{10mm} \bm\Omega=\Omega\bm e_3, \hspace{10mm} \bm B_0=B_0(x_2)\bm e_1, \hspace{10mm} \bm J_0=J_0\bm e_3,
\end{equation}
and
\begin{equation}
\se
\frac{\mathrm d B_0}{\mathrm d x_2}=-\mu J_0.
\end{equation}
\eme

In the third case $B_0=B_0(x_2,x_3)$, considered in section \ref{Bx-23}, we  
assume that $\partial^2B_0/\partial^2x_2=\partial^2B_0/\partial^2x_3=0$. For a large-scale magnetic field, we make the plausible assumption that
its curvature (second-order derivative) is negligible compared to its slope (first-order derivative). Therefore the third case is a superposition of case 1 and case 2.

In all the three cases of the basic state, the Lorentz force $\bm J_0\times\bm B_0$ is curl-free and can be balanced by the pressure gradient $-\bm\nabla P_0$, and therefore the Navier-Stokes equation in the basic state is reduced to $-\bm\nabla P_0+\bm J_0\times\bm B_0=\bm 0.$

\section{Case of $B_0=B_0(x_3)$\label{Bx-2}}
Substituting (\ref{decomposition}a-d) into the N-S equation (\ref{Navier-Stokes}), we derive the linearised perturbation equation for $\bm u$:
\begin{equation}
\label{Navier-Stokes-u}
\frac{\partial\bm u}{\partial t}=-\frac{1}{\rho}\bm\nabla p+\nu\nabla^2\bm u+2\bm u\times\bm\Omega+\frac{1}{\rho}\left(\bm j\times\bm B_0+\bm J_0\times\bm b\right).
\end{equation}
Taking the curl of (\ref{Navier-Stokes-u}) to eliminate the pressure gradient and employing the expressions (\ref{background1}c,d) for  $\bm B_0$ and $\bm J_0$, we obtain the vorticity equation,
\begin{equation}\label{vorticity}
\frac{\partial\bm\omega}{\partial t}=\nu\nabla^2\bm\omega+2\Omega\frac{\partial}{\partial x_3}\bm u+\frac{1}{\rho}\left(B_0\frac{\partial}{\partial x_1}\bm j-\mu J_0j_3\bm e_1-J_0\frac{\partial}{\partial x_2}\bm b\right).
\end{equation}
Similarly, from the magnetic induction equation (\ref{magnetic-induction}), we derive the linearised perturbation equation for $\bm b$:
\begin{equation}\label{induction}
\frac{\partial\bm b}{\partial t}=B_0\frac{\partial}{\partial x_1}\bm u-\mu J_0u_3\bm e_1+\eta\nabla^2\bm b.
\end{equation}
In the derivation of (\ref{vorticity}) and (\ref{induction}), the solenoidal properties $\bm\nabla\bm\cdot\bm B_0=\bm\nabla\bm\cdot\bm b=\bm\nabla\bm\cdot\bm J_0=\bm\nabla\bm\cdot\bm j=0$ have been employed.

Usually the large-scale background field $\bm B_0(x_3)$ varies much more smoothly than the small-scale perturbations, and accordingly its associated electric current $\bm J_0$ is very weak. We now make an approximation that the length scale of the background field is much longer than that of the perturbations, namely
\begin{equation}
\frac{1}{B_0}\frac{\mathrm d B_0}{\mathrm d x_3}=\frac{\mu J_0}{B_0}\sim\frac{1}{L}\ll k,
\end{equation}
where $L$ is the length scale of the background field and $k$ is the wavelength of perturbations. This is the so-called short-wavelength approximation. With this approximation $B_0$ in the perturbation equations (\ref{vorticity}) and (\ref{induction}) can be treated as uniform, but its gradient or $J_0$, though very weak, plays an important role in driving the instability. Certainly, this approximation is not rigorous, but it is plausible in the geophysical and astrophysical context, for which the large-scale azimuthal field (background field $\bm B_0$) varies much more smoothly than the small-scale field (field perturbation $\bm b$) induced by small-scale flow (flow perturbation $\bm u$). Therefore by considering a scale separation between the large-scale background magnetic field and the small-scale perturbed fluid velocity and magnetic field we can apply the short-wavelength approximation to our linear stability analysis. The key point of this approximation is that the background field varies smoothly and its associated electric current is weak but this weak current provides energy for instability. Moreover, this approximation can be validated through normal mode analysis in a plane layer geometry (rotating MHD flow between two infinite plates), in which the perturbation amplitudes are a function of $x_3$ and the velocity and magnetic boundary conditions might be crucial for the calculation of neutral stabilty curve. We do not attempt the normal mode analysis in this short paper.

Subject to the short-wavelength approximation, we assume that the perturbations have the form
\begin{equation}\label{wave}
\left(\bm u, \bm\omega, \bm b, \bm j\right)=\left(\hat{\bm u}, \hat{\bm\omega}, \hat{\bm b}, \hat{\bm j}\right)\exp\bigl[\mathrm i\left(\bm k\bm\cdot\bm x-ft\right)\bigr],
\end{equation}
where $\bm k=(k_1,k_2,k_3)$ is the wave vector and $f$ is the the frequency. Accordingly, we have the relationship for the amplitudes, $\hat{\bm\omega}=\mathrm i\bm k\times\hat{\bm u}$ and $\mu\hat{\bm j}=\mathrm i\bm k\times\hat{\bm b}$. Substituting the forms (\ref{wave}) into the perturbation magnetic induction equation (\ref{induction}) we obtain 
\begin{equation}\label{b}
\hat{\bm b}=\frac{\mu J_0 \hat u_3\bm e_1-\mathrm ik_1B_0\hat{\bm u}}{\mathrm if-\eta k^2},
\end{equation}
and accordingly
\begin{equation}\label{j}
\hat{\bm j}=\frac{1}{\mathrm if-\eta k^2}\left(\mathrm ik_3J_0\hat u_3\bm e_2-\mathrm ik_2J_0\hat u_3\bm e_3+k_1\frac{B_0}{\mu}\bm k\times\hat{\bm u}\right).
\end{equation}
Substituting (\ref{wave}), (\ref{b}) and (\ref{j}) into the perturbation vorticity equation (\ref{vorticity}) and noting the solenoidal property $\bm k\bm\cdot\hat{\bm u}=0$ of the velocity, we derive an equation for $\hat{\bm u}$:
\begin{equation}\label{curl1}
\left(\mathrm if-\nu k^2\right)\mathrm i\bm k\times\hat{\bm u}+2\mathrm ik_3\Omega\hat{\bm u}+\frac{k_1^2}{\rho\left(\mathrm if-\eta k^2\right)}\left[B_0J_0\left(\hat u_1\bm e_2-\hat u_2\bm e_1\right)+\frac{B_0^2}{\mu}\left(\mathrm i\bm k\times\hat{\bm u}\right)\right]=\bm 0.
\end{equation}
We then perform the operation $\mathrm i\bm k\times$ on (\ref{curl1}) to obtain
\begin{equation}\label{curl2}
\left(\mathrm if-\nu k^2\right)k^2\hat{\bm u}-2k_3\Omega\bm k\times\hat{\bm u}+\frac{k_1^2}{\rho\left(\mathrm if-\eta k^2\right)}\left(-\mathrm ik_3B_0J_0+\frac{B_0^2}{\mu}k^2\right)\hat{\bm u}=\bm 0.
\end{equation}
Again, we repeat the operation $\bm k\times$ on (\ref{curl2}) to obtain
\begin{equation}\label{curl3}
\left(\mathrm if-\nu k^2\right)k^2\bm k\times\hat{\bm u}+2k_3\Omega k^2\hat{\bm u}+\frac{k_1^2}{\rho\left(\mathrm if-\eta k^2\right)}\left(-\mathrm ik_3B_0J_0+\frac{B_0^2}{\mu}k^2\right)\bm k\times\hat{\bm u}=\bm 0.
\end{equation}
We combine (\ref{curl2}) with (\ref{curl3}) to eliminate $\bm k\times\hat{\bm u}$ and so derive a quadratic equation for $f$:
\begin{equation}\label{dispersion}
\left(\mathrm if-\nu k^2\right)k^2+\frac{k_1^2}{\rho\left(\mathrm if-\eta k^2\right)}\left(-\mathrm ik_3B_0J_0+\frac{B_0^2}{\mu}k^2\right)=\pm 2\mathrm ik_3k\Omega.
\end{equation}
The two solutions to (\ref{dispersion}) are
\begin{subequations}
\label{disp}
\begin{align}
f&=\frac{1}{2}\left[\pm f_\Omega-\mathrm i\left(\eta+\nu\right)k^2+\sqrt{\left[f_\Omega\pm \mathrm i\left(\eta-\nu\right)k^2\right]^2+4f_B^2\left(1-\mathrm i\frac{k_3\mu J_0}{k^2B_0}\right)}\,\,\right]\label{f1}\\
\intertext{and}
f&=\frac{1}{2}\left[\pm f_\Omega-\mathrm i\left(\eta+\nu\right)k^2-\sqrt{\left[f_\Omega\pm \mathrm i\left(\eta-\nu\right)k^2\right]^2+4f_B^2\left(1-\mathrm i\frac{k_3\mu J_0}{k^2B_0}\right)}\,\,\right],\label{f2}
\end{align}
\end{subequations}
where 
\bme
\begin{equation}
f_\Omega=\frac{2k_3\Omega}{k}
\qquad\qquad\mbox{and}\qquad\qquad
f_B=\frac{k_1B_0}{\sqrt{\rho\mu}}
\end{equation}
\eme
are the inertial wave frequency and Alfv\'en wave frequency respectively.
Equations (\ref{f1},b)
are the dispersion relationships for the perturbations. For
instability the imaginary part of $f$ should be positive, i.e. $f_I>0$. 
If the diffusivities are neglected ($\nu=\eta=0$) in order to eliminate damping, and the electric current vanishes ($J_0=0$) so removing a source of instability, then the two waves are obtained, i.e. the fast inertial wave and the slow magnetostrophic wave \citep[discussed in detail by][]{moffatt}.

We now make some simplifications. The simplest case is non-rotating ($\Omega=0$) inviscid ($\nu=0$) and perfectly conducting ($\eta=0$) MHD. For that case
(\ref{dispersion}) reduces to
\begin{equation}
f^2=\frac{k_1^2}{\rho k^2}\left(\frac{B_0^2}{\mu}k^2-\mathrm ik_3B_0J_0\right).
\end{equation}
The imaginary part shows that instability always occurs as long as $J_0$ is non-zero. With the short-wavelength approximation the ratio of the two terms in the brackets is
\begin{equation}
\frac{k_3B_0J_0}{B_0^2k^2/\mu}\sim\frac{k_3}{Lk^2}\ll\frac{k_3}{k}\leq1.
\end{equation}
Although the second term involving $J_0$ is very small compared to the first term, it cannot be neglected because the weak $J_0$ leads to the instability. This is the essence of the magnetic instability in our analysis.

We now consider the geophysical and astrophysical context. In the situation of $\nu\ll\eta$ which is usual in liquid metal (geophysical context) and plasma (astrophysical context), e.g. in the Earth's core the magnetic Prandtl number $Pm=\nu/\eta\approx10^{-6}$, we neglect viscosity ($\nu=0$). In the rapidly rotating system which is usual in the geophysical and astrophysical context, we employ the weak field approximation, namely
\begin{equation}\label{weakfield}
f_B\ll f_\Omega.
\end{equation}
We consider the Taylor series expansions of (\ref{f1},b) up to first order:
\begin{subequations}
\begin{align}
f_1&=+\left[\frac{f_B^2}{f_\Omega}\left(1-\mathrm i\frac{k_3\mu J_0}{k^2B_0}\right)-\frac{\eta^2k^4}{4f_\Omega}\right], \\[0.2em]
f_2&=-\left[\frac{f_B^2}{f_\Omega}\left(1-\mathrm i\frac{k_3\mu J_0}{k^2B_0}\right)-\frac{\eta^2k^4}{4f_\Omega}\right]-\mathrm i\eta k^2, \\[0.2em]
f_3&=+\left[f_\Omega+\frac{f_B^2}{f_\Omega}\left(1-\mathrm i\frac{k_3\mu J_0}{k^2B_0}\right)-\frac{\eta^2k^4}{4f_\Omega}\right], \\[0.2em]
f_4&=-\left[f_\Omega+\frac{f_B^2}{f_\Omega}\left(1-\mathrm i\frac{k_3\mu J_0}{k^2B_0}\right)-\frac{\eta^2k^4}{4f_\Omega}\right]-\mathrm i\eta k^2.
\end{align}
\end{subequations}
The imaginary part of $f_1$ and $f_3$ is
\begin{equation}
f_I=-\frac{k_1^2B_0J_0}{2\rho\Omega k},
\end{equation}
which indicates that the resistive rotating MHD flow becomes unstable if 
\begin{equation}\label{f13}
\Omega B_0J_0<0.
\end{equation}
For example, if $\Omega>0$ and $B_0>0$ but $J_0<0$ (or $dB_0/dx_3<0$) then flow is unstable. The imaginary part of $f_2$ and $f_4$ is
\begin{equation}
f_I=\frac{k_1^2B_0J_0}{2\rho\Omega k}-\eta k^2,
\end{equation}
which yields another criterion for the onset of instability,
\begin{equation}\label{f24}
\frac{k_1^2B_0J_0}{2\rho\eta\Omega k^3}>1.
\end{equation}
Equation (\ref{f24}) indicates that the resistive rotating MHD flow is unstable if the orientation of $\Omega$, $B_0$ and $J_0$ is such that $\Omega B_0J_0>0$ and the imposed field and current are sufficiently strong for the inequality (\ref{f24}) to be satisfied (although the weak field approximation (\ref{weakfield}) is still valid). In summary, when  $\nu\ll\eta$ and $f_B\ll f_\Omega$, the magnetic instability in rotating MHD occurs either in the orientation corresponding to $\Omega B_0J_0<0$ or in the orientation corresponding to $\Omega B_0J_0>0$ with the criterion (\ref{f24}).

\section{Case of $B_0=B_0(x_2)$ \label{Bx-3}}
In this section we consider the case of $B_0=B_0(x_2)$ in which the background field varies slowly in the radial direction. Many of derivations are similar to the case of $B_0=B_0(x_3)$ of section \ref{Bx-2} and so we do not show all the details.

The perturbed vorticity and magnetic induction equations are 
\begin{equation}
\label{vorticity-Bx-3}
\frac{\partial\bm\omega}{\partial t}=\nu\nabla^2\bm\omega+2\Omega\frac{\partial}{\partial x_3}\bm u+\frac{1}{\rho}\left(B_0\frac{\partial}{\partial x_1}\bm j+\mu J_0j_2\bm e_1-J_0\frac{\partial}{\partial x_3}\bm b\right)
\end{equation}
and
\begin{equation}
\label{ind-Bx-3}
\frac{\partial\bm b}{\partial t}=B_0\frac{\partial}{\partial x_1}\bm u+\mu J_0u_2\bm e_1+\eta\nabla^2\bm b,
\end{equation}
respectively. With the short-wavelength approximation, the perturbed magnetic field and electric current obtained from (\ref{ind-Bx-3}) are
\begin{equation}
\label{ind-b}
\hat{\bm b}=\frac{-\mu J_0 \hat u_2\bm e_1-\mathrm ik_1B_0\hat{\bm u}}{\mathrm if-\eta k^2}
\end{equation}
and
\begin{equation}
\label{ind-j}
\hat{\bm j}=\frac{1}{\mathrm if-\eta k^2}\left(-\mathrm ik_3J_0\hat u_2\bm e_2+\mathrm ik_2J_0\hat u_2\bm e_3+k_1\frac{B_0}{\mu}\bm k\times\hat{\bm u}\right)
\end{equation}
respectively. Substitution of (\ref{ind-b}) and (\ref{ind-j}) into the vorticity equation (\ref{vorticity-Bx-3}) yields
\begin{equation}
\label{ind-u}
\left(\mathrm if-\nu k^2\right)\mathrm i\bm k\times\hat{\bm u}+2\mathrm ik_3\Omega\hat{\bm u}+\frac{k_1^2}{\rho\left(\mathrm if-\eta k^2\right)}\left[B_0J_0\left(\hat u_1\bm e_3-\hat u_3\bm e_1\right)+\frac{B_0^2}{\mu}\left(\mathrm i\bm k\times\hat{\bm u}\right)\right]=\bm 0.
\end{equation}
As done in the last section, on applying the curl action ($\mathrm i\bm k\times$) and the double curl action ($\mathrm i\bm k\times(\mathrm i\bm k\times)$) to
(\ref{ind-u}),
we derive the quadratic equation
\begin{equation}
\label{dispersion-new}
\left(\mathrm if-\nu k^2\right)k^2+\frac{k_1^2}{\rho\left(\mathrm if-\eta k^2\right)}\left(\mathrm ik_2B_0J_0+\frac{B_0^2}{\mu}k^2\right)=\pm 2\mathrm ik_3k\Omega.
\end{equation}
The two solutions of (\ref{dispersion-new}) are
\begin{subequations}
\label{disp-new}
\begin{align}
f&=\frac{1}{2}\left[\pm f_\Omega-\mathrm i\left(\eta+\nu\right)k^2+\sqrt{\left[f_\Omega\pm \mathrm i\left(\eta-\nu\right)k^2\right]^2+4f_B^2\left(1+\mathrm i\frac{k_2\mu J_0}{k^2B_0}\right)}\,\,\right]
\intertext{and}
f&=\frac{1}{2}\left[\pm f_\Omega-\mathrm i\left(\eta+\nu\right)k^2-\sqrt{\left[f_\Omega\pm \mathrm i\left(\eta-\nu\right)k^2\right]^2+4f_B^2\left(1+\mathrm i\frac{k_2\mu J_0}{k^2B_0}\right)}\,\,\right].
\end{align}
\end{subequations}
A comparison of the dispersion relationships (\ref{disp}a,b) and (\ref{disp-new}a,b)
for the two cases $B_0=B_0(x_3)$ and $B_0=B_0(x_2)$ respectively
shows that (\ref{disp-new}a,b) is obtained from (\ref{disp}a,b) by simply
replacing the phase factor $-\mathrm ik_3$ with $+\mathrm ik_2$.

For non-rotating, ideal MHD, the dispersion relationship reduces to
\begin{equation}
f^2=\frac{k_1^2}{\rho k^2}\left(\frac{B_0^2}{\mu}k^2+\mathrm ik_2B_0J_0\right),
\end{equation}
which is consistent with the short-wavelength approximation in terms of the ratio of two terms in the brackets
\begin{equation}
\frac{k_2B_0J_0}{B_0^2k^2/\mu}\sim\frac{k_2}{Lk^2}\ll\frac{k_2}{k}\leq1.
\end{equation}
At a low $Pm$ and with the weak field approximation $f_B\ll f_\Omega$, we expand this dispersion relationship to the first order. The imaginary part of complex frequency has two expressions
\bme
\begin{equation}
+\frac{k_1^2k_2B_0J_0}{2\rho\Omega kk_3} \hspace{15mm}\text{and}\hspace{15mm} -\frac{k_1^2k_2B_0J_0}{2\rho\Omega kk_3}-\eta k^2.
\end{equation}
\eme
Accordingly the magnetic instability occurs either in the orientation such that
\bme
\begin{equation}
\se
\Omega B_0J_0k_2k_3>0
\end{equation} 
or in the orientation such that
\begin{equation}
\Omega B_0J_0k_2k_3<0 
\qquad\qquad\mbox{with}\qquad\qquad
\frac{k_1^2k_2B_0J_0}{2\rho\eta\Omega k^3k_3}<-1.
\end{equation}
\eme
Compared to the case of axial dependence of background field, in the case of radial dependence of background field, the signs of $k_2$ and $k_3$ should be taken into account, namely whether the wavevector of a plane wave is inward or outward in the radial direction and whether it is parallel or anti-parallel to the rotation axis are important for the onset of magnetic instability.

\section{Case of $B_0=B_0(x_2,x_3)$\label{Bx-23}}
As explained in the section \ref{B-S},
this $B_0=B_0(x_2,x_3)$  case is a superposition of the last two cases
$B_0=B_0(x_3)$ of section \ref{Bx-2} 
and $B_0=B_0(x_2)$ of section \ref{Bx-3}. 
Accordingly, we can readily give the results. 
In particular the dispersion relationship, that extends 
(\ref{disp}a,b) and (\ref{disp-new}a,b), is 
\begin{subequations}
\label{disp-newer}
\begin{align}
f&=\frac{1}{2}\left[\pm f_\Omega-\mathrm i\left(\eta+\nu\right)k^2+\sqrt{\left[f_\Omega\pm \mathrm i\left(\eta-\nu\right)k^2\right]^2+4f_B^2\left(1+\mathrm i\frac{(k_2-k_3)\mu J_0}{k^2B_0}\right)}\,\,\right]\\
\intertext{and}
f&=\frac{1}{2}\left[\pm f_\Omega-\mathrm i\left(\eta+\nu\right)k^2-\sqrt{\left[f_\Omega\pm \mathrm i\left(\eta-\nu\right)k^2\right]^2+4f_B^2\left(1+\mathrm i\frac{(k_2-k_3)\mu J_0}{k^2B_0}\right)}\,\,\right].
\end{align}
\end{subequations}
At a low $Pm$ and with the weak field approximation, the imaginary part of frequency is
\bme
\begin{equation}
+\frac{k_1^2(k_2-k_3)B_0J_0}{2\rho\Omega kk_3} \hspace{15mm}\text{and}\hspace{15mm} -\frac{k_1^2(k_2-k_3)B_0J_0}{2\rho\Omega kk_3}-\eta k^2.
\end{equation}
\eme
Therefore, the magnetic instability occurs either in the orientation such that
\bme
\begin{equation}
\se
\Omega B_0J_0(k_2-k_3)k_3>0
\end{equation}
or in the orientation such that
\begin{equation}
\Omega B_0J_0(k_2-k_3)k_3<0
\qquad\qquad\mbox{with}\qquad\qquad
\frac{k_1^2(k_2-k_3)B_0J_0}{2\rho\eta\Omega k^3k_3}<-1.
\end{equation}
\eme
Compared to the previous two cases $B_0=B_0(x_3)$ of section \ref{Bx-2} 
and $B_0=B_0(x_2)$ of section \ref{Bx-3}, 
the difference between $k_2$ and $k_3$ plays a role in the onset of magnetic instability in our third case $B_0=B_0(x_2,x_3)$ considered here.

\section{Summary \label{Summary}}
In this short paper, we have undertaken the local analysis for magnetic instability in rotating MHD, in which the weak imposed electric current drives the instability. We have adopted a local Cartesian coordinate system and have taken advatage of scale separation in the short-wavelength approximation. Three cases of imposed field have been studied, namely the axial dependence in section \ref{Bx-2} , the radial dependence in section \ref{Bx-3} and the mixed case in section \ref{Bx-23}. The orientation of rotation $\bm\Omega$, imposed magnetic field $\bm B_0$ and imposed electric current $\bm J_0$ is crucial to this instability. Moreover, in the second case ${\bm B}_0=B_0(x_2){\bm e}_1$ of section \ref{Bx-3} the direction of wavevector is important, and in the third case ${\bm B}_0=B_0(x_2,x_3){\bm e}_1$  of section \ref{Bx-23} the difference between radial and axial wavenumbers is important for the onset of magnetic instability.

\section*{Acknowledgments}
The motivation of this calculation was illuminated by the talk with Prof. G\"unther R\"udiger about the paper \citep{jones} during my visit to Astrophysical Institute Potsdam. An anonymous referee pointed out that the radial dependence of background field would be more interesting to astrophysicists. I am financially supported by the project SPP1488 of the program PlanetMag of Deutsche Forschungsgemeinschaft (DFG).

\bibliographystyle{gGAF}
\bibliography{Wei-edit}

\end{document}